\documentstyle[aps,preprint]{revtex}

\newcommand{\pade}{Pad\'{e}} 
\newcommand{\btf}{$\beta$~function}

\newcommand{\mh}{\mbox{$M_H$}}
\newcommand{\gf}{\mbox{$G_F$}} 

\newcommand{\padef}[2]{{\mbox{$\beta\,[{#1},\,{#2}]$}}}

\begin{document}
\draft
\tighten

\title{Perturbative reliability of the Higgs-boson coupling in the standard
electroweak model} 

\author{Loyal Durand\thanks{Electronic address: ldurand@theory2.physics.wisc.edu}                
\ and Gregory Jaczko\thanks{Electronic address: 
jaczko@theory1.physics.wisc.edu}}                
\address{Department of Physics, University of Wisconsin-Madison,\\ 
1150 University Avenue, Madison, WI 53706} 

\date{\today} 
\maketitle

\begin{abstract}
We apply \pade\ summation to the $\beta(\lambda)$ function for the 
quartic Higgs coupling $\lambda$ in the standard electroweak model.
We use the $\beta$ function calculated to five loops in the 
minimal subtraction scheme to demonstrate the improvement resulting
from the summation, and then apply the method to the more physical
on-mass-shell renormalization scheme where $\beta$ is known to three loops.  
We conclude that the OMS $\beta$ function and the running coupling 
$\lambda(\mu)$ are reliably known over the range of energies and Higgs-boson
masses of current interest.
\end{abstract}


\section{INTRODUCTION} 
In the standard model of elementary particle physics, the $SU{\left( 2
  \right)} \times U{\left( 1 \right)}$ symmetry is spontaneously
broken to a residual $U{\left(1\right)}_{EM}$, generating mass for the $W^\pm$
and $Z$ gauge bosons and the matter fields.  A possible cause for the
symmetry breaking is the presence of an additional scalar field, the
Higgs field. Although there is as yet no experimental evidence for 
the expected Higgs boson, we can still explore the
implications of this symmetry breaking mechanism using radiative corrections
to standard-model processes.  For
example, the condition that perturbative calculations be reliable
provides a theoretical upper bound on the mass \mh\ of a weakly interacting 
Higgs boson 
\cite{dicus,lqt,chan,marciano,dawson,passarino,vayonakis,djl,ldurand,durand2,kurt96}.

In a recent paper \cite{kurt96}, 
Nierste and Riesselmann analyzed one- and two-scale processes involving
the Higgs field with a particular emphasis on the running of the
quartic Higgs coupling $\lambda(\mu)$. They assessed the
reliability of perturbation theory using two criteria: the relative
difference of physical quantities calculated in different
renormalization schemes; and the dependence of $\lambda$ on the 
renormalization scale $\mu$.
If perturbation theory is to be reliable, 
the choices of the renormalization 
scheme and scale should not be important for physical
quantities. 
To determine $\lambda(\mu)$ in their analysis,
Nierste and Riesselmann integrated the renormalization group
equation using the three-loop $\beta$ function, and solved the resulting
equation for $\lambda$ iteratively using four different approximation schemes.
The solutions differed significantly for large values of the coupling
or mass scale, and determined one constraint on \mh\ in a perturbative theory. 
This uncertainty in $\lambda(\mu)$ carries over to physical quantities
such as scattering amplitudes and again affects the ranges of \mh\ and $\mu$
over which perturbative calculations are reliable.

We show here that is is possible to explore the regime of large coupling
without the ambiguities that arise from the direct iterative solution for the
coupling.  We approach the problem by emphasizing the $\beta$ function
$\beta(\lambda)$, and show that it can be determined reliably to rather
large values of $\lambda$ by using \pade\ approximates 
\cite{baker,BendOrsz} to sum the
perturbation series for $\beta(\lambda)$
\cite{yang}. Integration of the renormalization 
group equation then gives an implicit equation for $\lambda(\mu)$ that can be
inverted numerically. The results can be used
to study the the validity of perturbation theory for 
scattering amplitudes in the region of large Higgs-boson masses 
and high energies where the running coupling $\lambda(\mu)$ 
is the natural renormalization group expansion parameter. We will not
pursue those applications here as a number of authors 
\cite{marciano,dawson,passarino,vayonakis,djl,ldurand,durand2,kurt96,ghinculov} have considered them in detail. 

We first investigate the \pade\ approach in Sec.\ \ref{sec:pade} using the
the results for $\beta(\lambda)$ in the minimal subtraction (MS) 
renormalization scheme for which the perturbation series for $\beta$
is known to five loops \cite{ms4loop,ms5loop}. 
After establishing the effectiveness of the \pade\
approach, we apply it in Sec.\ \ref{sec:applications} to the more physical 
on-mass-shell (OMS)
renormalization scheme where $\beta$ is only known to three loops
\cite{kurt96,luscher}. We
find that \pade\ summation of the series apparently
gives a reliable result for $\beta(\lambda)$
for quite large values of the coupling,
$\lambda\leq 10$, and conclude, after inversion of the renormalization group
expression, that $\lambda(\mu)$ is known reliably in the OMS scheme for
$\mu\leq 4$ TeV for $\mh\leq 800$ GeV. The region in which $\lambda(\mu)$ is 
known well extends to very large mass scales if \mh\ is sufficiently small, 
for example, to $10^{17}$ GeV for $\mh\leq 155$ GeV.

\section{PAD\'{E} SUMMATION OF THE $\beta$ FUNCTION}
\label{sec:pade}

\subsection{Preliminary considerations}
\label{subsec:prelim}

In the following, we deal with the quartic Higgs-boson coupling $\lambda$ 
defined at tree level in terms of \mh\ and the electroweak vacuum expectation
value $v=246$ GeV or the Fermi coupling \gf\ by $\lambda=\mh^2/2v^2=\gf\mh^2/\sqrt{2}$. 
We will work in the interesting 
limit of large Higgs-boson masses, corresponding
to the limit of large quartic couplings, and neglect the effects of couplings 
with fermions. 

The running coupling $\lambda(\mu)$ is defined as the solution of the
renormalization group equation
\begin{equation}
\label{RGeqn}
\mu\,\frac{d\lambda(\mu)}{d\mu}=\beta(\lambda)
\end{equation}
at the energy scale $\mu$. The function $\beta(\lambda)$ is given in 
perturbation theory as a power series in $\lambda$,
\begin{eqnarray}
\beta(\lambda) &=& \frac{\lambda^2}{16\pi^2}\,\sum_{n=0}\,\beta_n\left(
\frac{\lambda}{16\pi^2}\right)^n \label{beta_series}\\
&=& \beta_0\,\frac{\lambda^2}{16\pi^2}\,\left(1+\sum_{n=1}\,B_n\lambda^n
\right). \label{B_series}
\end{eqnarray}
The coefficients $\beta_n$ are renormalization-scheme dependent beyond two
loops. They are known through three loops in the on-mass-shell renormalization
scheme \cite{kurt96,luscher},
\begin{equation}
{\rm OMS:}\quad \beta_0=24,\quad \beta_1=-312,\quad \beta_2=4238.23,
\label{beta_n_oms}
\end{equation}
and to five loops in the minimal subtraction scheme \cite{ms4loop,ms5loop},
\begin{eqnarray}
{\rm MS:}\quad &\beta_0=24, \quad \beta_1=-312, \quad \beta_2=12022.7
\nonumber\\[1ex]
& \beta_3=-690759, \quad \beta_4=4.91261\times 10^7. \label{beta_n_ms}
\end{eqnarray}
Alternatively, the coefficients $B_n$ are given by
\begin{equation}
{\rm OMS:} \quad B_0=1,\quad B_1=-0.082323,\quad B_2=0.0070816, \label{B_n_oms}
\end{equation}
\begin{eqnarray}
{\rm MS:}\quad & B_0=1,\quad B_1=-0.082323,\quad B_2=0.020089 \nonumber\\[1ex]
& B_3=-0.0073090, \quad B_4=0.0032917. \label{B_n_ms}
\end{eqnarray}

To determine the running coupling, one must integrate the renormalization 
group  equation, Eq.\ (\ref{RGeqn}), and solve the implicit equation
\begin{equation}
\ln\frac{\mu}{\mu_0}=
        \int^{\lambda \left(\mu \right)}_{\lambda \left( \mu_0 \right)} 
        \frac{dx}{\beta(x)}.
\label{beta:int}
\end{equation}
This equation determines $\lambda(\mu)$ in terms of the initial and final mass
scales $\mu_0$ and $\mu$ and the initial value of the coupling at the scale 
$\mu_0$, defined as $\lambda_0=\lambda(\mu_0)$.  
Different, typically iterative, methods of solution lead to 
different results for $\lambda(\mu)$, with the differences increasing
for large vales of \mh\ or $\lambda_0$ and for $\mu\gg\mu_0$ \cite{kurt96}. 
Since the $\beta$ function is only
known to finite order, the only constraint on this standard approach is that 
the different solutions satisfy the renormalization group equation,
Eq.\ (\ref{beta:int}), to that order.  However, the
resulting ambiguities for large values of \mh\  can compromise tests
of the reliability of perturbation theory, and the determination of limits
on \mh\ in a weakly interacting theory. It is therefore useful to approach
the problem differently, and concentrate on the $\beta$ function itself. If
$\beta(\lambda)$ is known accurately for some range of $\lambda$, the 
integral in Eq.\ (\ref{beta:int}) will also be accurately determined, and
the equation can be inverted numerically to find $\lambda(\mu)$ in that
region.

\subsection{Pad\'{e} summation and $\beta(\lambda)$}
\label{subsec:pade}
 
\pade\ approximates \cite{baker,BendOrsz} give a very useful way of summing or extrapolating
series for which only a finite number of terms are known. 
The $[N,\,M]$ \pade\ approximate for a function 
$f(z)$ defined by a truncated power series
\begin{equation}
f(z)=\sum_{j=0}^m\,c_jz^j + O(z^{m+1})
\end{equation}
is a ratio of two polynomials,
\begin{equation}
P[N,\,M](z)\equiv\frac{\textstyle{\sum_{n=0}^{N}\,a_nz^n}}
{\textstyle{\sum_{n=0}^{M}\,b_nz^n}},\quad b_0=1,\quad 
N+M=m.\label{pade_define}
\end{equation}
The coefficients $a_n,\,b_n$ are determined uniquely by the
requirement that the series expansion of 
$P[N,\,M](z)$ agree term-by-term with the series for $f(z)$ through terms of 
order $z^m$. 

The sequence of \pade\ approximates $P[N,\,M]$ is known to converge to
$f(z)$ as $N,\,M\rightarrow\infty$ with $N-M$ fixed for large classes
of functions \cite{baker,BendOrsz}, but the approximates can also give useful
and rapidly convergent asymptotic approximations for finite $N$ and $M$
even if the sequence and the original series for $f(z)$ do not converge \cite{BendOrsz}. 

In the present case, the function in question is $\beta(\lambda)$, known
perturbatively to orders $\lambda^4$ and $\lambda^6$, that is, to three
and five loops, in the OMS and MS renormalization schemes, respectively. 
The perturbation series for $\beta$ is not expected to converge, 
but a \pade\ summation
of the series may still be useful for $\lambda$ not too large. Because the
perturbative expansion of $\beta(\lambda)$ starts at order $\lambda^2$, we
will extract the leading power explicitly, redefine the \pade\ coefficients,
and define the $[N,\,M]$ approximate for the n-loop $\beta$ function as
\footnote{\pade\ summation of $\beta$ was considered by Yang and Ni 
\cite{yang}, but without applications to the present problem. Those authors
did not extract the overall factor $\lambda^2$, so use a different 
labeling of the approximates, and miss the diagonal approximates used 
here.} 

\begin{equation}
\beta[N,\,M]=\beta_0\frac{\lambda^2}{16\pi^2}\,\frac{1+a_1\lambda+
a_2\lambda^2+\cdots+a_N\lambda^N}
{1+b_1\lambda+b_2\lambda^2+\cdots+b_M\lambda^M},\quad N+M=n-1.
\label{eq:P_for_beta} 
\end{equation} 
Note that the approximates \padef{n-1}{0} are just the perturbation series
for $\beta$ carried to n loops.

The series for $\beta(\lambda)$ defined by Eq.\ (\ref{B_series}) are
alternating series in which the ratios of coefficients $B_{n+1}/B_n$ 
change only slowly in either OMS or MS
renormalization in the range in which the $B$'s are known. This suggests that
the diagonal approximates \padef{N}{N} with $M=N$ or the
subdiagonal approximates with $M=N+1$ may be particularly effective in
estimating the series. In the case of OMS renormalization, the $\beta$'s are
known only to three loops, so $M+N\leq 2$. The possible choices are then
\padef{1}{1} or \padef{0}{2} if we use all the three-loop information,
or \padef{0}{1} if the perturbation series is truncated at two loops. 
\padef{2}{0} and \padef{1}{0} are just the three- and two-loop
perturbation series. In the case of MS renormalization, $\beta$ is known to 
five loops, $M+N\leq 4$, and we will consider the approximates
\padef{1}{2} at the four-loop level, and \padef{2}{2} at five loops, keeping
$M=N$ or $M=N+1$. The additional five-loop approximates
\padef{1}{3}, \padef{3}{1}, and \padef{0}{4} are members of sequences 
two or more steps off the diagonal. These are not expected to converge as
rapidly as the sequences we consider. The coefficients
$a_j$, $b_j$ for these approximates are given in appendix A.

\subsection{Tests of \pade\ summation using MS renormalization}
\label{subsec:pade_tests}

The fact that the perturbation series for $\beta$ is known to five loops
gives us the opportunity to test the \pade\ summation procedure using
known results. Having established its reliability, we will the apply the
method in Sec.~\ref{sec:applications} to the more physical OMS renormalization
scheme in which the connection between $\lambda$ and \mh\ is known.

\subsubsection{Convergence of the Pad\'{e} sequence}
\label{subsubsec:convergence}

Based upon the general convergence properties of \pade\
approximates and the alternating character of the series at hand, 
we expect the sequence \padef{1}{1}, \padef{1}{2}
and \padef{2}{2} to converge as we progress from three to five loops.
We plot these approximates in Fig.~\ref{fig:ms5} 
to demonstrate that convergence.  
The convergence of the \pade\
sequence is, in fact,
relatively fast.  For low values of $\lambda$ there is
excellent agreement.  Even for $\lambda=10$, \padef{1}{1} and
\padef{1}{2} differ by $<$ 10 \%  with the diagonal five-loop approximate 
\padef{2}{2} lying roughly halfway between the other two. 
We interpret the agreement and the pattern of convergence as strong
evidence for the effectiveness of the \padef{N}{N} sequence
in summing the series for $\beta(\lambda)$, and conclude that it is
unlikely that $\beta$ would be found to differ significantly from 
\padef{2}{2} in the region shown if higher-loop contributions were
calculated.

In Fig.\ \ref{fig:n0ms}, we look at the problem from the point of view 
of the purely perturbative approach, and show the sequence of the 
N-loop perturbation series \padef{N-1}{0} for $\beta$. 
This is not a sequence in which $N$
and $M$ increase together with the difference $N-M$ fixed, so the
standard results on \pade\ convergence do not apply.
The convergence of the sequence is very slow as shown in the figure,
with large differences between successive terms already present for
$\lambda \simeq 3$.  For
comparison, we also show the three- and five-loop diagonal approximates
\padef{1}{1} and \padef{2}{2}.  These forms 
interpolate the perturbative sequence very well, eliminating
the dominance of the last term in the series for $\lambda$ large.   
Since \padef{1}{1}, \padef{2}{2}, and the four-loop approximate \padef{1}{2} 
differ from each other by less than
5\% for $\lambda<10$, all are effective in extrapolating the perturbation 
series. We conclude, in particular, that the three-loop approximate
\padef{1}{1} already gives a reliable extrapolation 
for $\beta(\lambda)$, with uncertainties of only a few percent, out to
$\lambda\sim 10$, far beyond the range in which the five-loop perturbation 
series is reliable.

\subsubsection{Estimates of unknown coefficients}
\label{subsubsec:coefficients}

 \pade\ approximates often converge to the limit function faster than
the power series used to construct them.  In that case, the
terms in the expansion of a \pade\ approximate beyond the matched
order may give reasonable estimates for the unknown higher-order
coefficients in the power series.  As a simple test of this expectation
in the present case, we can expand the three- and four-loop approximates
\padef{1}{1} and \padef{1}{2} to one order higher in $\lambda$ than the
finite power series used to construct them, and compare the new coefficient
with the known four- and five-loop results. Thus, the expansion
\begin{equation}
   \padef{1}{1} = \frac{\lambda^2}{16\pi^2}\,\beta_0\,\left[ 1+B_1\lambda +
    B_2\lambda^2 + (B_2^2/B_1)\lambda^3 + (B_2^3/B_1^2)\lambda^4 +\cdots
    \right] 
 \end{equation}
gives the estimates
\begin{equation}
  \label{eq:B3_est}
  B_3^\prime  \equiv B_2^2/B_1, \qquad B_4^\prime \equiv B_2^3/B_1^2,
\end{equation}
for the four- and five-loop coefficients $B_3$ and $B_4$, results equivalent to
\begin{equation}
\label{eq:b3_est} 
 \beta_3^\prime \equiv\beta_2^2/\beta_1 = -463286,\qquad \beta_4^\prime
\equiv \beta_2^3/\beta_1^2 = 1.785\times 10^7.
\end{equation}
The actual four- and five-loop results are
\begin{equation}
  \label{eq:b3_act}
  \beta_3 = -690759, \qquad \beta_4=4.913\times 10^7.
\end{equation}
The estimates of $\beta_3$ and $\beta_4$ from the three-loop
coefficients
are therefore about 0.67 and 0.36 of the actual coefficients.

In the case of \padef{1}{2}, we can estimate only $\beta_4$, with the
result
\begin{equation}
 \label{eq:b4_est}
 B_4^\prime= -\left(B_3^2-2B_1B_2B_3+B_2^3\right)/\left(B_1^2-B_2\right).
\end{equation}
This estimate gives $\beta_4^\prime=3.48\times 10^7$, and a ratio
$\beta_4^\prime/\beta_4=0.71$. 

The estimates for the first missing terms in the perturbation series
are too small in both of the cases considered. We can understand this 
result qualitatively as resulting from the averaging of an alternating
series by the approximates, with the corresponding tendency to avoid large
higher coefficients in the expansion. We will use this observation below.

The effects of incorrect estimates of $B_3$ on the approximate \padef{1}{2}
are shown in Fig.\ \ref{fig:b3'}. In these calculations, we have taken
$B_3$ as five- and ten times the estimated value, and calculated \padef{1}{2}
using the new value as input. The result is a $<10$\% change in
$\beta$ for $\lambda<10$ despite the very large values of the new coefficient.

\subsection{The running coupling $\lambda(\mu)$ in MS renormalization}
\label{subsec:MSrunning}

The effect of the uncertainty in $\beta(\lambda)$ on the running of 
$\lambda(\mu)$ can be studied by integrating the renormalization group
equation, Eq.\ (\ref{RGeqn}), and solving numerically for $\lambda$
as a function of its initial value $\lambda_0$ and the ratio of energy
scales $\mu/\mu_0$.\footnote{In the case of MS renormalization, $\lambda$ is
connected only indirectly to the physical pole mass of the Higgs boson, 
so we cannot state the results in terms of $\mu$ and \mh\ without
using a separate calculation of the self-energy function.}
We have done this calculation using the approximates
\padef{1}{1} and \padef{1}{2}, choosing initial values $\lambda_0=1,\,3,\,5$.
The results are shown in Fig.\ \ref{fig:MSrunning}. The result for 
the optimum five-loop approximate, \padef{2}{2}, lies near the center of the 
shaded regions in that figure, as would be expected from the comparison
of the approximates in Fig.\ \ref{fig:ms5}. We believe the estimated range
of uncertainty is quite generous given the rapid convergence of the
sequence shown there toward \padef{2}{2}. 

The range of uncertainty in $\lambda(\mu)$ at fixed $\mu/\mu_0$ is quite
small for $\lambda_0=1,\,3$ over the entire range shown, $\mu/\mu_0\leq 6$. 
The uncertainty is larger for $\lambda_0=5$, roughly 16\%, at $\mu/\mu_0=3$, 
but even then the boundary curves differ from the curve for \padef{2}{2}
by $<8$\%.

The rather small effect of uncertainties in $\beta$ on $\lambda(\mu)$ can be
understood rather simply. The renormalization group equation involves
$1/\beta$ rather than $\beta$. The prefactor $\lambda^2$ in the \pade\
expression in Eq.\ (\ref{eq:P_for_beta}) leads to a rapid decrease in the
integrand, and the value of the integral is determined mainly by the region
near $\lambda_0$, the lower endpoint of the integration.  For $\lambda_0$ 
small, $\beta$ is well determined in the most important region, and the
uncertainty in the integral is small. The uncertainty in the integral,
hence the uncertainty in $\lambda(\mu)$, becomes 
large only for renormalization group evolution  away from a large
starting value for $\lambda_0$.

\section{APPLICATIONS: RANGES OF RELIABILITY OF $\beta(\lambda)$ and 
$\lambda(\mu)$ IN OMS RENORMALIZATION}
\label{sec:applications} 

\subsection{\pade\ approximates for $\beta(\lambda)_{OMS}$}
\label{subsec_beta_OMS} 

Having tested the use of \pade\ approximates in the MS scheme, we
consider the implications of \pade\ summation for the OMS scheme.
The most significant difference is the limited order, three loops, 
to which the perturbation series for $\beta$ is known. 
We are therefore restricted to two approximates that use the full 
information available, the diagonal approximate \padef{1}{1} and
the subdiagonal approximate \padef{0}{2}.  We can also use \padef{0}{1}
at the two-loop level.
Based upon the convergence of the \pade\ sequence demonstrated for the MS 
scheme, and the apparent reduction in the size of the coefficients in the OMS
scheme\footnote{The known value of $\beta_2$ in the OMS scheme is smaller
than that in the MS scheme by roughly a factor of three
\cite{kurt96,luscher}. Nierste and
Riesselmann \cite{kurt96} have found similar reductions in the coefficients in 
the expansion of physical amplitudes. We assume that the reductions in the
size of the coefficients persist at higher orders.}, 
we will assume that these approximates again provide an accurate estimate
for the \btf, with the diagonal approximate probably the most reliable.

To determine the range of $\lambda$ for which the \btf\ is reliable, we first
considered the differences among the three-loop functions \padef{1}{1} and
\padef{0}{2}, and the two-loop function
\padef{0}{1}. These approximates can barely be distinguished over 
the range of $\lambda$ shown in Fig.~\ref{fig:p11p02} with the scale used 
there, so only \padef{1}{1} is shown. This agreement is the result of the
nearly geometric growth of the first coefficients in the perturbation series.
The three-loop approximates \padef{1}{1} and \padef{0}{2} continue to agree 
well to much larger values of $\lambda$. 
While one is tempted on this basis to conclude that the
OMS \btf\ is reliably known for $\lambda\leq 10$, the range of current
interest, the geometric character of the low-order perturbation series
may well be accidental. We have therefore 
attempted to estimate a wider range of uncertainty in the \btf\
in a different way by supposing, in agreement with the results of the
MS analysis, that the coefficient $B_3^\prime$ estimated by expanding
\padef{1}{1} is too small, and constructing a new ``four-loop'' approximate \padef{1}{2} using a greatly increased value of $B_3$.  
The result obtained using $B_3=5B_3^\prime$ 
is shown in Fig.~\ref{fig:p11p02}. The change in the extrapolation
of the perturbation series is quite small, with a difference
of less than 2\% between \padef{1}{1} and \padef{1}{1} for $\lambda<10$. 
We also show the perturbation series for the \btf, \padef{2}{0}, in 
Fig.~\ref{fig:p11p02} for comparison.

\subsection{The running coupling $\lambda(\mu)$}
\label{sec:running_coupling}

In the OMS renormalization scheme, the parameter
$\lambda$ is defined by the relation
$\lambda=\gf\mh^2/\sqrt{2}$ to all orders in perturbation theory 
\cite{ldurand,ldkr}. We will choose the starting value $\lambda_0$ of the 
running coupling $\lambda(\mu)$ to have this value.
What remains to be decided is the energy scale $\mu_0$
at which this relation should be taken to hold. The natural energy
scale would appear to be $\mu_0=\mh$. However, other choices have been
made. Thus, in an early investigation, Sirlin and Zucchini \cite{SirZuch}
calculated the one-loop corrections to the four-point Higgs-boson scattering
amplitude and defined the parameters in the theory so that large
electromagnetic effects appear only in such standard relations as that 
between \gf\ and the muon decay rate. With this definition, the high mass 
limit of the four-point function gives \cite{SirZuch}
\begin{equation}
h(\mu)=\lambda_0\left[1+\frac{\lambda_0 }{16\pi^2}\left(
24\ln{\frac{\mu}{\mh}}+25-3\sqrt{3}\pi \right)\right].
\label{eq:sirlin}  
\end{equation}

The logarithm in the expression above is just that which appears in 
the expansion of the one-loop expression for $\lambda(\mu)$,
\begin{equation}
\lambda(\mu)=\lambda_0\left(1-\beta_0\frac{\lambda_0}{16\pi^2}\,\ln\frac{\mu}
{\mu_0}\right)^{-1}
\label{eq:one_loop_lambda}
\end{equation}
for $\beta_0=24$ and $\mu_0=\mh$. The ambiguity in the choice of $\mu_0$ 
is in the treatment of the remaining constants in Eq.\ (\ref{eq:sirlin}). 
These have been incorporated in the running coupling
by some authors\cite{SirZuch,djl,ldurand} 
by redefining $\mu_0$ as $\mu_0=\mh\exp[(-25+3\sqrt{3}\pi)/24]$. 
However, the constants do not
appear naturally in the expression for the four-point function at two loops
\cite{ldurand}. It is probably most reasonable, therefore, to treat them as
separate ``radiative corrections'' and write $h(\mu)$ to one loop as
$h(\mu)=\lambda(\mu)[1+\delta]$, with $\lambda(\mu)$ the one-loop
running coupling defined above, and $\delta$ incorporating the remaining
scale-independent corrections.

This question has been studied in more detail by Nierste and Riesselmann
\cite{kurt96}, who showed that the convergence of the perturbation series
was improved for several physical amplitudes by adopting the natural
scale $\mu_0=\mh$ instead of the choice noted above. 
They note, furthermore, that in order
to cancel large logarithmic terms in the perturbative
result when one considers two-scale physical processes such as scattering,
the scale $\mu$ must be related to the energy scale of the interaction 
by $\mu=\sqrt{s}$ \cite{kurt96}. We will follow Nierste and Riesselmann
and make the definite, physically motivated choices 
$\mu_0=\mh$ and $\mu=\sqrt{s}$
in the  following analysis. This specification amounts as already noted to
a definite specification of the ``radiative corrections'' in perturbatively
calculated amplitudes once the couplings are expressed in terms of 
$\lambda(\mu)$. 

With $\lambda_0$ and $\mu_0$ specified, and the range of reliability of
the \btf\ established, it is straightforward to integrate the renormalization
group equation and invert the result numerically to obtain $\lambda(\mu)$.
The uncertainty in $\lambda(\mu)$ can be specified in terms of that in $\beta$.
With this procedure, no further uncertainties such as those illustrated
in \cite{kurt96} are introduced. It is not necessary, for example, to obtain 
the solution of the renormalization group equation as a series in $\lambda_0$, 
in an iterative approximation. We note in this connection that the ``naive''
and ``consistent'' forms for $\lambda(\mu)$ given in \cite{kurt96}
correspond respectively to the approximates \padef{N}{0}, the perturbation
series for $\beta$, and \padef{0}{N}, the series obtained by expanding
$\/\beta$. Neither sequence is expected to converge well with increasing $N$.

Our results for $\lambda(\mu)$ are shown in Fig.~\ref{fig:lmaxvmu} for
$\mh=500$ and 800 GeV and $\mu_0\leq\mu\leq 4$ TeV. We find
for $\mh=500$ GeV that all \pade\
approximates, including the perturbation series, agree very for $\mu<5$ TeV, 
a region in which $\lambda_0<5$. The
residual uncertainty in $\lambda(\mu)$ is small enough not to affect
perturbative results for physical processes.

Different \pade\ approximates also give very similar extrapolations
for $\lambda(\mu)$ for $\mh=800$ GeV, even when the predicted value of
$\beta_3$ is changed by a large factor. The only significant deviation 
involves the perturbation series \padef{2}{0} which we do not believe
is reliable on the basis of our earlier investigation. Even if we restrict
the range of $\lambda$  in which we take $\beta$ as reliable to $\lambda<10$
as in Fig.~\ref{fig:p11p02}, the result for $\lambda(\mu)$ remains
reliable for $\mu<2$ TeV, a value well into the energy region of interest
for experiments at the Large Hadron Collider at CERN.

The rapid growth of $\lambda(\mu)$ for the perturbative
approximate \padef{2}{0} in Fig.~\ref{fig:lmaxvmu} is the result of an
Landau pole at $\mu=2339$ GeV.
A pole can appear in $\lambda(\mu)$ if the
integral of $1/\beta$ converges for $\lambda\rightarrow\infty$, with
the position of the pole in $\mu$ determined by the condition
\begin{equation}
        \ln\frac{\mu}{\mu_0}=\lim_{\lambda \rightarrow \infty}
                        \int_{\lambda_0}^\lambda \frac{d\lambda}{\beta}.
\label{eq:Landau}
\end{equation}
No pole can actually appear when the integration is restricted 
to the finite range of $\lambda$ in which $\beta$ is known reliably, 
but the likely presence of 
a pole would be indicated by very rapid growth of $\lambda(\mu)$ with 
increasing $\mu$ in that region. In the present case, 
there is no reason to expect the perturbation series \padef{2}{0}
to be accurate for $\lambda$ large. The results in 
Fig.~\ref{fig:p11p02} indicate, in fact, 
that the perturbative approximation
begins to fail badly for $\lambda\approx 5$, while the starting point for the 
evolution of $\lambda(\mu)$ shown in Fig.~\ref{fig:lmaxvmu} is 
at $\lambda=5.3$ for $\mh=800$ GeV. The remaining approximates do not lead to 
poles in the region shown. 

\subsection{Conclusions}
\label{subsec:conclusions}
We have shown that \pade\ summation of the $\beta$ function 
improves the reliability of $\beta$ and the running quartic Higgs coupling.
The method gives a best estimate for $\beta$, and removes much of the
uncertainty associated with different determinations of $\lambda(\mu)$
at the three-loop level \cite{kurt96}. 

We have tested the \pade\ method using the
\btf\ in the MS renormalization scheme, where $\beta$ is known to five
loops in the perturbation expansion. The test results suggest rapid
convergence of the diagonal and subdiagonal \pade\ sequences. Our applications
are to the more physical OMS renormalization scheme, where  the first
scheme-dependent coefficient in the OMS expansion is significantly smaller
than in the MS expansion. This more rapid apparent convergence is reflected
in the excellent agreement among the leading \pade\ approximates for
$\beta_{OMS}$ even for rather large values of the first unknown
coefficient, $\beta_3$.

Application of the \pade\ method to the three-loop results for the 
OMS \btf\ leads to a running coupling that appears to be quite reliable
in the region studied, namely a Higgs-boson mass $\mh\leq 800$ GeV and 
a mass scale $\mu\leq 4$ TeV. We conclude that uncertainties in $\lambda(\mu)$
are not an important source of uncertainty in perturbative results for
physical scattering and decay amplitudes in this interesting region.

\acknowledgments
This work was supported in part by the U.S. Department of Energy under 
Grant No.\ DE-FG02-95ER40896.
One of the authors (LD) would like to thank the Aspen Center for Physics
for its hospitality while parts of this work were done.

\appendix

\section{PAD\'{E} APPROXIMATES}
\label{app:pade_forms}

The coefficients of the \pade\ forms used in our analysis are given below. 
We will state the results in terms of the coefficients  
$B_n=(\beta_n/\beta_0)/(16\pi^2)^n$.
 
At two loops, $N+M=1$ and we have only the truncated perturbation series 
\padef{1}{0} and the approximate \padef{0}{1} with
\begin{equation}
\padef{0}{1}:\qquad  b_1=-B_1
\end{equation}

At three loops, $N+M=2$ and
we have the new approximants \padef{1}{1} and \padef{0}{2}.
The coefficients are given by: 
  \begin{eqnarray}         
   \padef{1}{1}:\qquad  a_1 &=& (B_1^2-B_2)/B_1,\label{eq:p11}\\
    b_1 &=& -B_2/B_1, \nonumber \\
   &&\nonumber\\
   \padef{0}{2}:\qquad b_1 &=& -B_1,\label{eq:p02}\\
    b_2 &=& B_1^2-B_2.\nonumber   
  \end{eqnarray}

At five loops, $N+M=4$ and we will consider the new approximants \padef{3}{1},
\padef{2}{2}, \padef{1}{3} and \padef{0}{4}. The coefficients are given by: 
  \begin{eqnarray}
  \padef{3}{1}:\qquad a_1 &=& (B_1B_3-B_4)/B_3, \label{eq:p31} \\
  a_2 &=& (B_2B_3-B_1B_4)/B_3, \nonumber \\
  a_3 &=& (B_3^2-B_2B_4)/B_3 ,\nonumber \\
  b_1 &=& -B_4/B_3 , \nonumber \\
  && \nonumber \\
  \padef{2}{2}:\qquad a_1 &=& (B_1B_2^2-B_1^2B_3+B_1B_4-B_2B_3)/A_{22},
    \label{eq:p22}\\
  a_2 &=& (B_2^3-2B_1B_2B_3+B_1^2B_4+B_3^2-B_2B_4)/A_{22}, \nonumber\\
  b_1 &=& (B_1B_4-B_2B_3)/A_{22}, \nonumber \\
  b_2 &=& (B_3^2-B_2B_4)/A_{22}, \nonumber\\
  A_{22} &=& B_2^2-B_1B_3, \nonumber \\
  &&\nonumber \\
  \padef{1}{3}:\qquad a_1 &=&( B_1^4-3B_1^2B_2+2B_1B_3+B_2^2-B_4)/A_{13}, 
   \label{eq:p13} \\
  b_1 &=& (-B_1^2B_2+B_2^2+B_1B_3-B_4)/A_{13}, \nonumber \\
  b_2 &=& (B_1B_2^2-B_2B_3-B_1^2B_3+B_1B_4)/A_{13}, \nonumber \\
  b_3 &=& (2B_1B_2B_3-B_2^3-B_3^2+B_2B_4-B_1^2B_4)/A_{13}, \nonumber \\
  A_{13} &=& B_1^3-2B_1B_2+B_3. \nonumber
 \end{eqnarray}

We will also consider the approximates \padef{1}{2}, the subdiagonal
approximate for the four-loop expansion. The coefficients in this case are:
\begin{eqnarray}
  \padef{1}{2}:\qquad a_1 &=& (B_1^3-2B_1B_2+B_3)/(B_1^2-B_2),
   \label{eq:p12} \\
  b_1 &=&  (B_3-B_1B_2)/(B_1^2-B_2), \nonumber \\
  b_2 &=& (B_2^2-B_1B_3)/(B_1^2-B_2). \nonumber
\end{eqnarray}

\section{ANALYTIC RESULTS}
\label{app:results}
The \pade\ approximates we have used are all integrable analytically.  
We will give only the results needed in our investigation of the
OMS renormalization scheme:

\begin{eqnarray}
\padef{0}{1}:\quad \frac{\beta_0}{16\pi^2}\, \int^\lambda
  \frac{d\lambda}{\padef{0}{1}}&=& -\frac{1}{\lambda}-B_1\ln{\lambda}, \\
&&\nonumber \\
\padef{2}{0}:\quad \frac{\beta_0}{16\pi^2}\, \int^\lambda
  \frac{d\lambda}{\padef{2}{0}}
  &=& -\frac{1}{\lambda} 
  -B_1\ln{\lambda} +\frac{1}{2}B_1
  \ln\left(1+B_1\lambda+B_2\lambda^2\right) \nonumber \\
  && +\frac{B_1^2-2 B_2}{\sqrt{4B_2-B_1^2}}
  \arctan{\frac{B_1+2B_2\lambda}{\sqrt{4B_2-B_1^2}}},\\
&&\nonumber \\
\padef{1}{1}:\quad \frac{\beta_0}{16\pi^2}\,\int^\lambda \frac{d\lambda}{\padef{1}{1}}
  &=&
-\frac{1}{\lambda}-B_1\ln{\lambda}\nonumber\\
&& +B_1\ln{\left(1+ \frac{B_1^2-B_2}{B_1}\lambda\right)},\\
  && \nonumber \\
\padef{0}{2}:\quad  \frac{\beta_0}{16\pi^2}\,\int^\lambda \frac{d\lambda}{\padef{0}{2}}
  &=& -\frac{1}{\lambda}
  -B_1\ln{\lambda}+\left(B_1^2-B_2\right)\lambda.
\end{eqnarray}
These expressions are to be equated to $(\beta_0/16\pi^2)\ln(\mu/\mu_0)$.

\newpage

\begin{figure}
\caption{ A comparison of the sequence of five-loop diagonal and 
subdiagonal \pade\ approximates for $\beta(\lambda)$ in the MS
renormalization scheme. Note that alternate approximates are too large
or too small, and that the sequence converges rapidly with the final
result presumably in the band between \padef{1}{2} and \padef{2}{2}.}
\label{fig:ms5}
\end{figure}

\begin{figure}
\caption{Demonstration of the slow convergence of successive 
perturbative approximations
to the \btf\ toward the diagonal \pade\ approximate \padef{2}{2} 
for MS renormalization.}
\label{fig:n0ms}
\end{figure}

\begin{figure} 
\caption{
Plots of the \pade\ approximate \padef{1}{2} to the MS \btf\ using the
actual value of the four-loop coefficient $\beta_3$ and values five and ten
times the estimate obtained from the three-loop approximate \padef{1}{1}.}
\label{fig:b3'}
\end{figure}

\begin{figure}  
\caption{
Plots showing the running of $\lambda(\mu)$ as a function of the ratio of
scales $\mu/\mu_0$ for different initial choices of $\lambda_0$ in the MS 
renormalization scheme. The differences between the curves
obtained using the \pade\ approximates \padef{1}{1} and \padef{1}{2}
corresponding to three- and four-loop summations of $\beta$ indicates the
range of uncertainty in the result. The curves for the five-loop \btf\
lie near the center of the band of uncertainty.}
\label{fig:MSrunning}
\end{figure}

\begin{figure}
\caption{Plots of the two- and three-loop \pade\ approximates \padef{0}{1}
and \padef{1}{1} for $\beta$ in the OMS scheme. The function \padef{1}{2}
obtained using a coefficient $\beta_3$ five times as large as that estimated
from \padef{1}{1} is shown to indicate a range of uncertainty. The 
three-loop perturbation series \padef{2}{0} is shown for comparison.}
\label{fig:p11p02}
\end{figure}

\begin{figure}
  \caption{Plots showing the running of $\lambda(\mu)$ 
in the OMS renormalization scheme as 
a function of the mass scale $\mu$ for different initial choices of 
the Higgs-boson mass \mh\ . The differences between the curves
obtained using the three-loop \pade\ approximate \padef{1}{1} and the  
function \padef{1}{2} obtained using a coefficient $\beta_3$ five times as 
large as that estimated from \padef{1}{1} is shown to indicate a range of 
uncertainty. The perturbative result for $\beta$ given by \padef{2}{0} has
a Landau pole at $\mu=2.3$ TeV for $\mh=800$ GeV, but is not reliable and is
included only to illustrate the effects of a nearby pole.
 }
  \label{fig:lmaxvmu}
\end{figure}

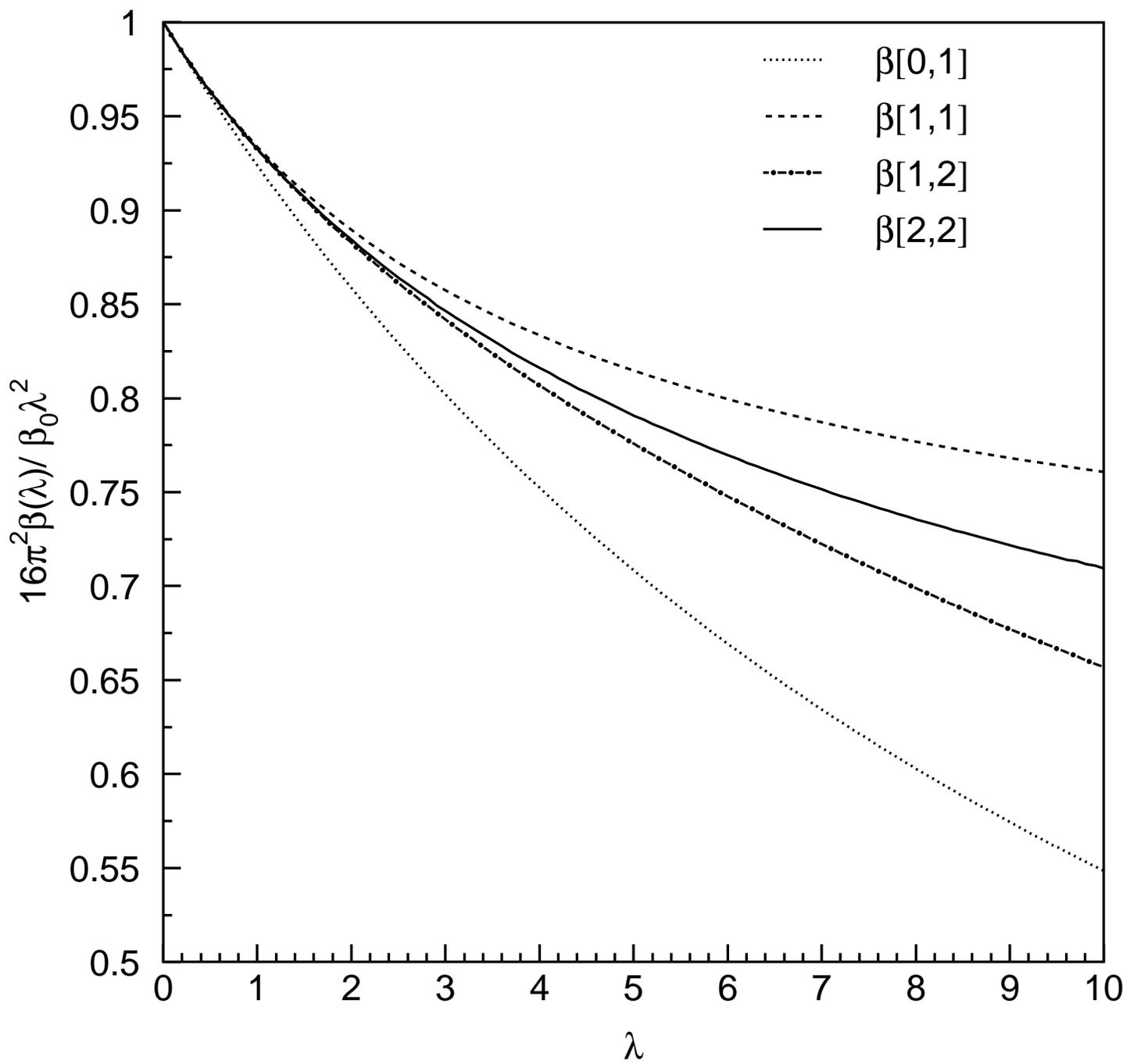
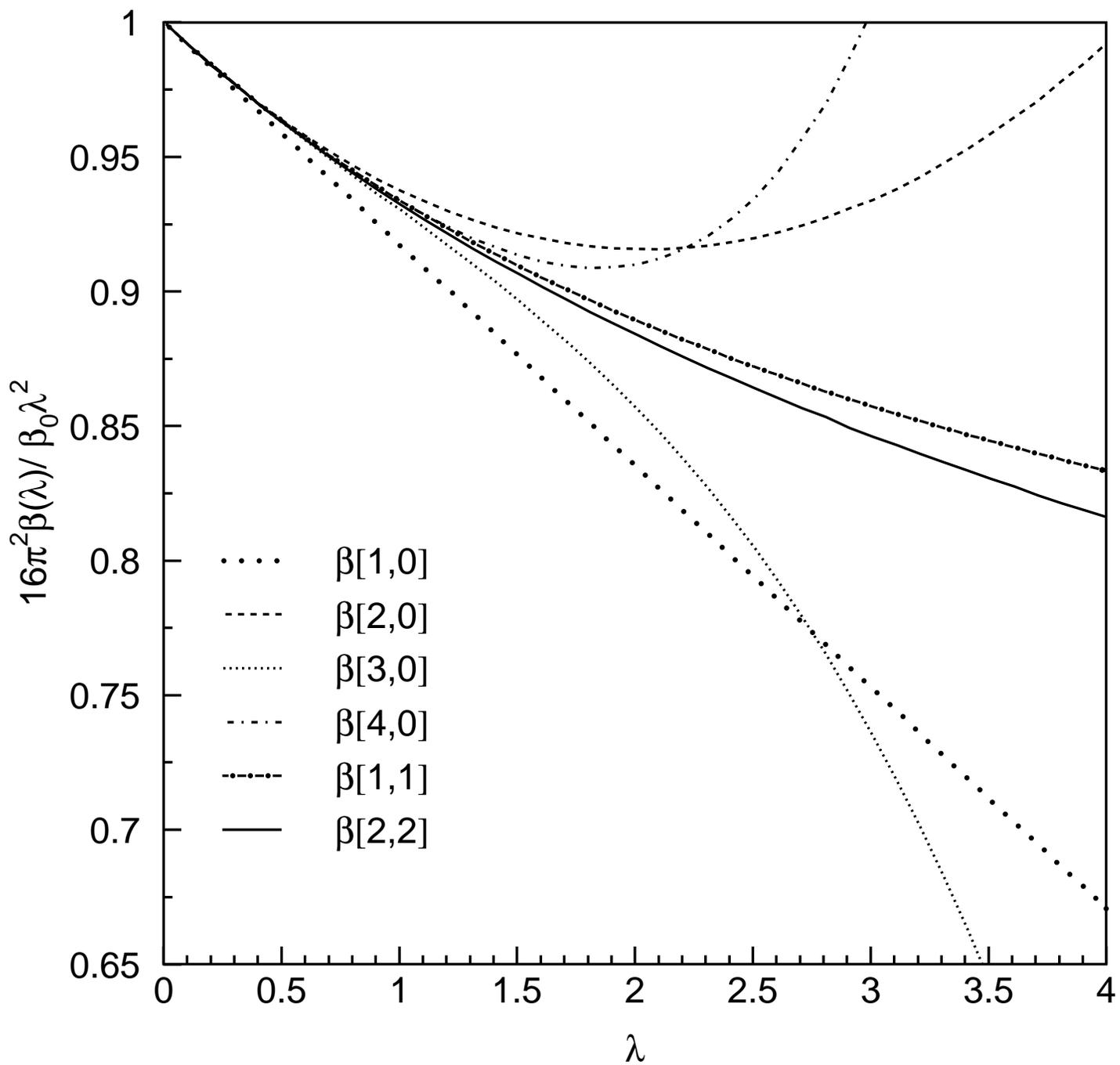
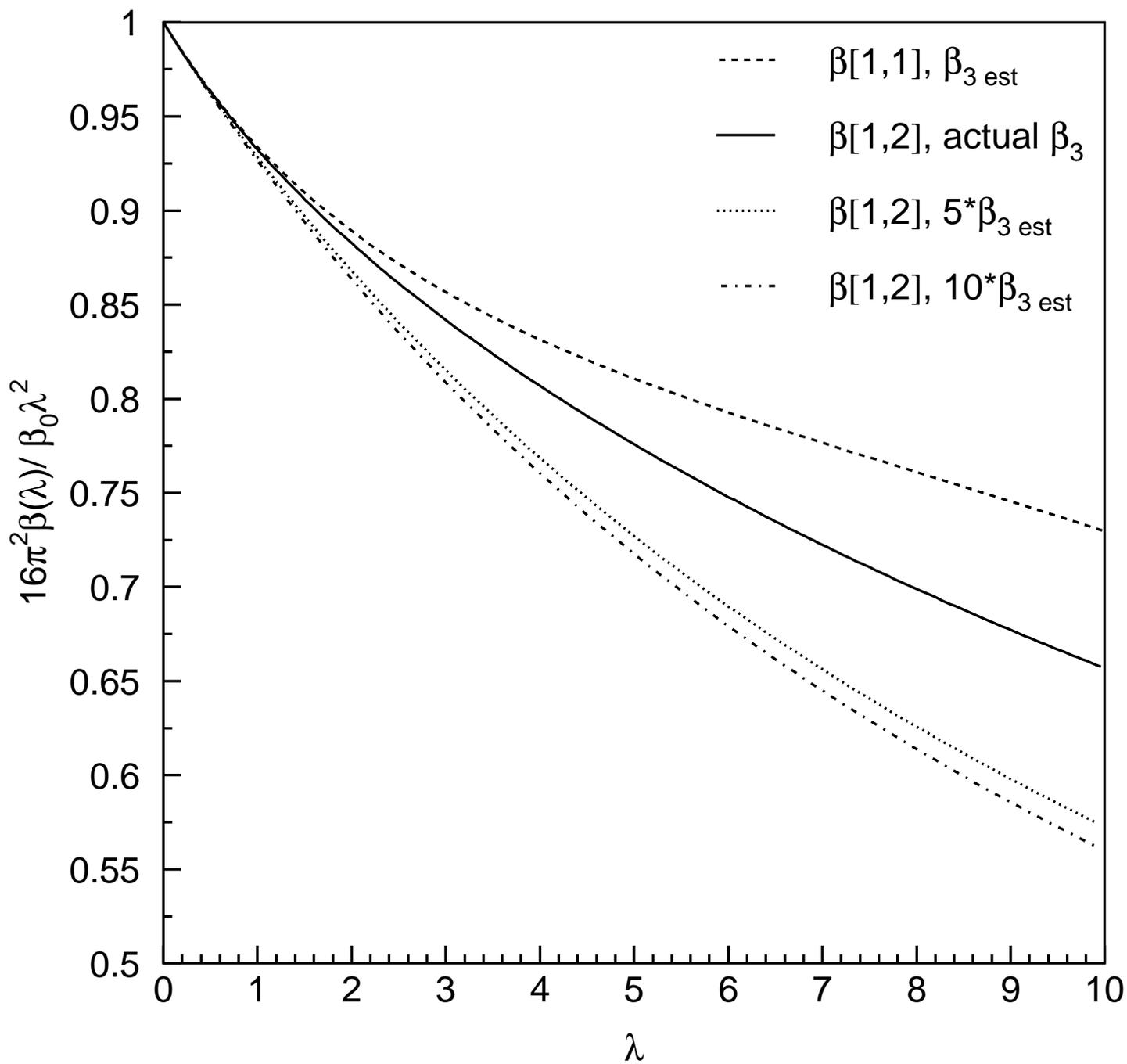
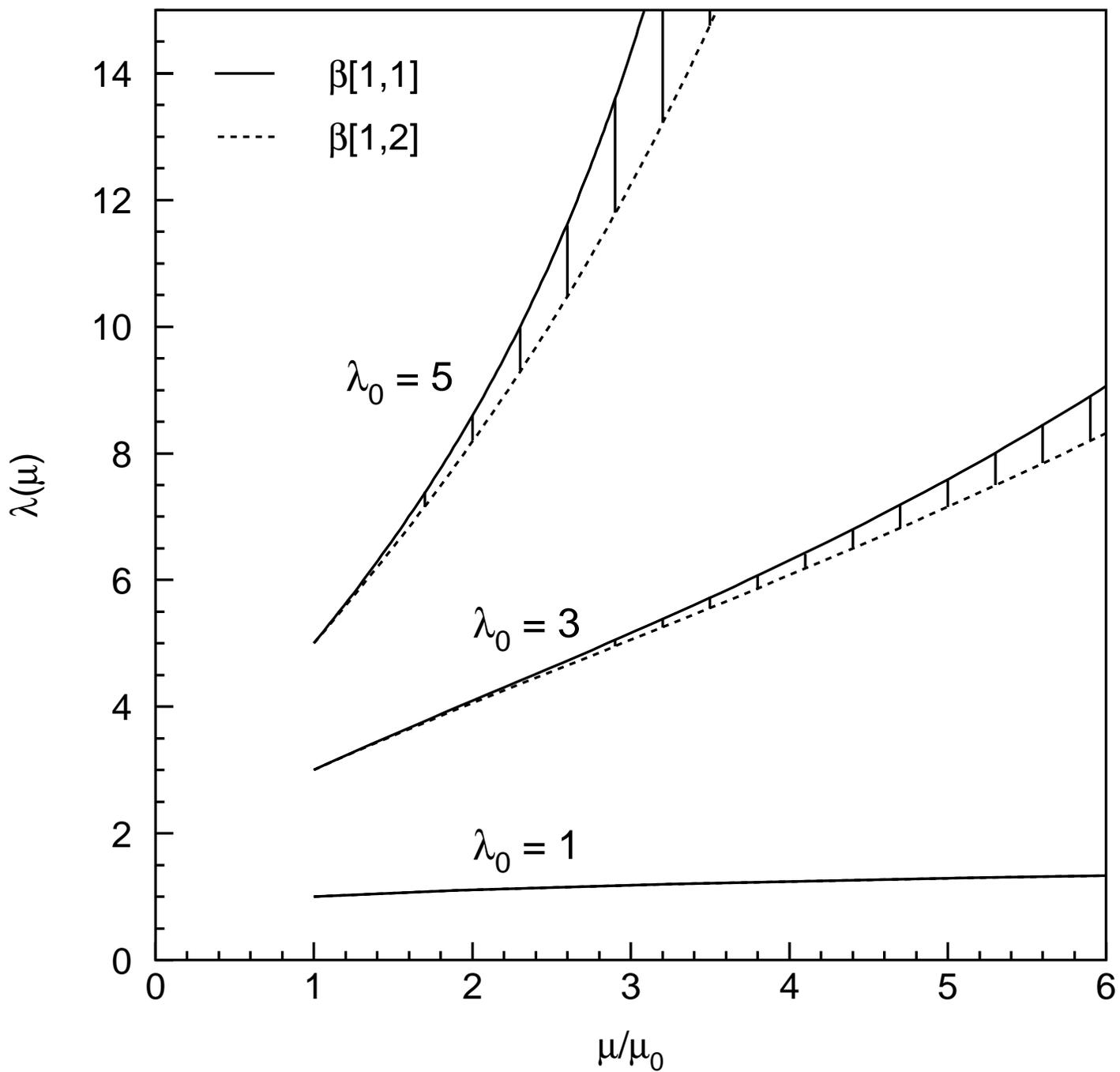
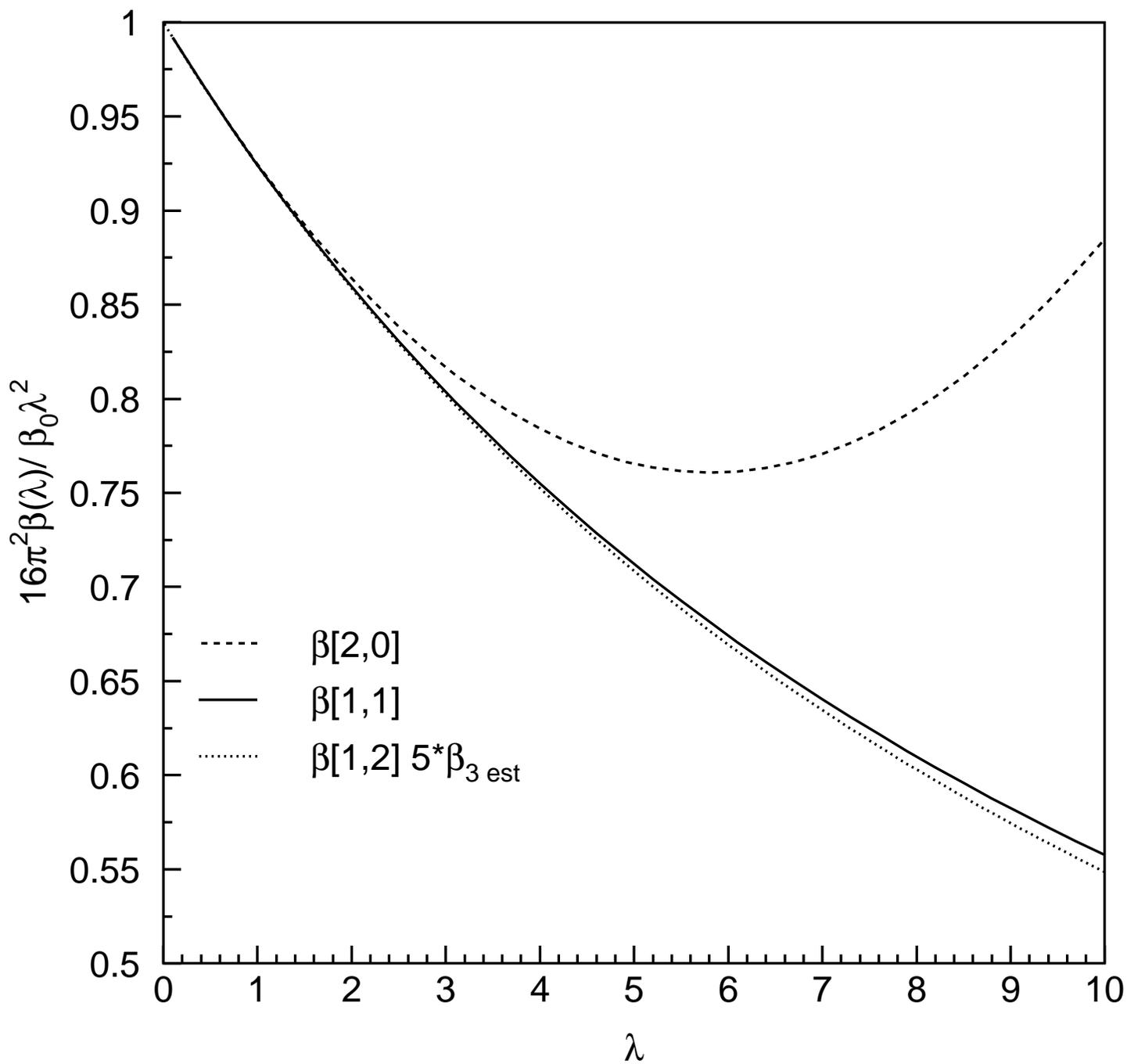
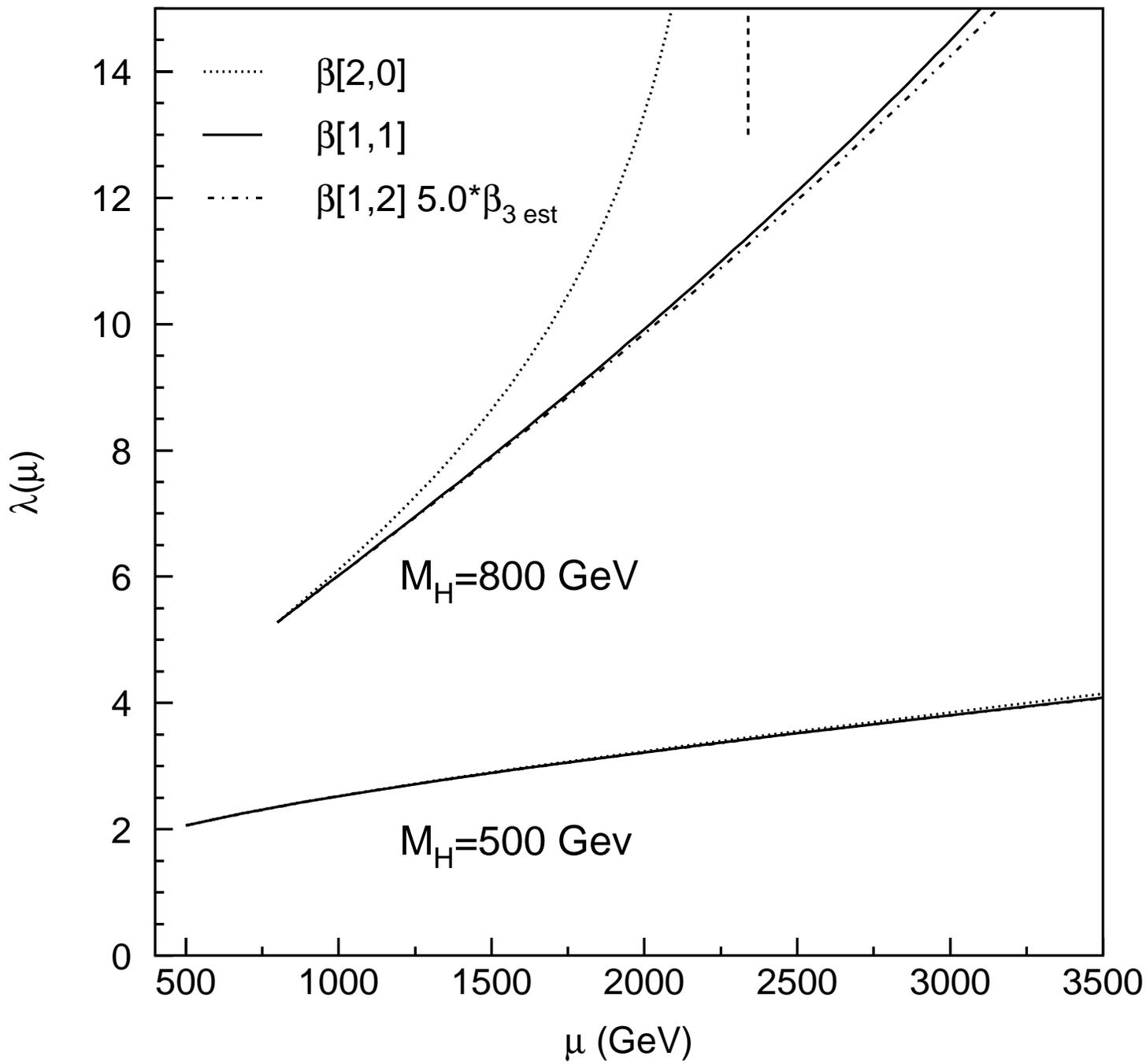
\end{document}